\documentclass[aps,prc,twocolumn,showpacs,superscriptaddress,floatfix]{revtex4}
\usepackage{graphicx}
\usepackage{dcolumn}
\usepackage{amsfonts}
\usepackage{amsmath}

\begin{document}

\title{Microscopic Calculation of Pre-Compound Excitation Energies\\ for Heavy-Ion Collisions}


\author{A.S. Umar}
\affiliation{Department of Physics and Astronomy, Vanderbilt University, Nashville, Tennessee 37235, USA}
\author{V.E. Oberacker}
\affiliation{Department of Physics and Astronomy, Vanderbilt University, Nashville, Tennessee 37235, USA}
\author{J.A. Maruhn}
\affiliation{Institut f\"ur Theoretische Physik, Goethe-Universit\"at, 60438 Frankfurt am Main, Germany}
\author{P.-G. Reinhard}
\affiliation{Institut f\"ur Theoretische Physik, Universit\"at Erlangen, D-91054 Erlangen, Germany}

\date{\today}


\begin{abstract}
We introduce a microscopic approach for calculating the excitation energies of systems formed
during heavy-ion collisions. The method is based on time-dependent Hartree-Fock (TDHF)
theory and allows the study of the excitation energy as a function of time or ion-ion
separation distance. We discuss how this excitation energy is related to the estimate
of the excitation energy using the reaction $Q$-value, as well as its implications for
dinuclear pre-compound systems formed during heavy-ion collisions.
\end{abstract}
\maketitle

During a heavy-ion collision the initial kinetic energy of the two nuclei is
gradually converted into internal excitation. This excitation may result in
exciting various types of collective modes or intrinsic excitations that
lead to the heating of the combined nuclear system. The excitation of
projectile-like and target-like fragments in deep-inelastic heavy-ion
collisions is a well known manifestation of this phenomenon, which has
been studied theoretically and experimentally~\cite{TS92}. On the other hand, for a
dinuclear pre-compound system formed during a heavy-ion collision, the mode of decay
may critically depend on the excitation energy of the system.
Examples include collisions that may be candidates for the formation of
superheavy elements  in hot or cold fusion reactions~\cite{Ho02,Og07}.
Excitation energy is also an important ingredient for the stochastic mean-field approach to nuclear 
dynamics, which deals with fluctuations of collective motion in addition to the
average evolution~\cite{Ayik08,Ayik09}.
While exclusive measurements of excitation energy may be possible for
equilibrated systems (e.g. compound nuclei formed in 
complete fusion or  fragments produced in deep-inelastic
collisions), the intermediate states formed during a collision have a short life time and are not expected
to be fully equilibrated thus making the measurement as well as the
interpretation very difficult~\cite{Br90}. For all these reasons it is desirable to
develop a dynamical approach for calculating the excitation energy of the
system as it evolves in time.

It is generally acknowledged that the TDHF theory provides a
useful foundation for a fully microscopic many-body theory of low-energy heavy-ion reactions
\cite{Ne82,DDKS}. While the long-time evolution in TDHF theory may not be very
reliable, recent three-dimensional TDHF calculations with no symmetry assumptions
and using modern Skyrme forces have shown to accurately reproduce phenomena
determined by the initial stages of the heavy-ion dynamics~\cite{GM08,UO09a,UOM08}.
Recently we have developed the density-constrained TDHF (DC-TDHF) method~\cite{UO06b},
which is based on the generalization of the density constraint method developed
earlier~\cite{CR85}. We have shown that using the DC-TDHF method
ion-ion potential barriers can be accurately produced~\cite{UO06d,UO08a,UO09b} as these
calculations also depend on early stages of the ion-ion dynamics. In addition, one-body energy
dissipation extracted from TDHF for low-energy fusion reactions was found to be in agreement with the
friction coefficients based on the linear response theory as well as those in models where
the dissipation was specifically adjusted to describe experiments~\cite{WL09}. All of these
new results suggest that TDHF dynamics provide a good description of the early stages of
heavy-ion collisions.

In this manuscript we outline a microscopic approach for calculating excitation energies of composite
or dinuclear systems formed during heavy-ion collisions. The goal of the approach is to
divide the TDHF motion into a collective and intrinsic part. The major assumption in
achieving this goal is to assume that the collective part is primarily determined by
the density $\rho(\mathbf{r},t)$ and the current $\mathbf{j}(\mathbf{r},t)$. Consequently,
the excitation energy can be formally written as
\begin{equation}
E^{*}(t)=E_{TDHF}-E_{coll}\left(\rho(t),\mathbf{j}(t)\right)\;,
\label{eq:1}
\end{equation}
where $E_{TDHF}$ is the total energy of the dynamical system, which is a conserved quantity,
and $E_{coll}$ represents the
collective energy of the system. In the next step we break up the collective energy into two parts
\begin{equation}
E_{coll}\left(t\right)= E_{kin}\left(\rho(t),\mathbf{j}(t)\right) + E_{DC}\left(\rho(t)\right)\;,
\end{equation}
where $E_{kin}$ represents the kinetic part and is given by
\begin{equation}
E_{kin}\left(\rho(t),\mathbf{j}(t)\right)=\frac{m}{2}\int\;{\rm d}^{3}r\;\mathbf{j}^2(t)/\rho(t)\;,
\end{equation}
which is asymptotically equivalent to the kinetic energy of the
relative motion, $\frac{1}{2}\mu\dot{R}^2$, where $\mu$ is the
reduced mass and $R(t)$ is the ion-ion separation distance.
The energy $E_{DC}$ is the lowest-energy state of all possible
TDHF states with the same density and is required to have zero excitation
energy. This state is found by using the density-constraint method~\cite{CR85,US85}, which
minimizes the energy while holding the instantaneous TDHF density constant.
We have previously shown~\cite{UO06b} that $E_{DC}$ equals the ion-ion
potential, $V(R)$, shifted by the binding energies of the participating nuclei,
which allows us to write
\begin{equation}
E_{coll}(t)= E_{kin}(\rho(t),\mathbf{j}(t)) + V(R(t)) + E_{A_1} + E_{A_2}\;,
\end{equation}
where $E_{A_1}$ and $E_{A_2}$ denote the Hartree-Fock energies calculated for the
two nuclei using the same effective interaction. The dynamics of the ion-ion separation
$R(t)$ can be extracted from an unrestricted TDHF run. Using $E^{*}(t)$ and $R(t)$, we can deduce
the excitation energy as a function of the distance parameter, $E^{*}(R)$.

The computation of the excitation energy as outlined above is numerically very intensive,
primarily due to the density-constraint calculation.
Calculations were done in 3-D geometry and using the full Skyrme force (SLy4)~\cite{CB98}
without the center-of-mass correction as described in
Ref.~\cite{UO06}. The numerical accuracy of the static binding energies and the deviation
from the point Coulomb energy in the initial state of the collision dynamics is on the order
of $50-200$~keV. We have performed density constraint calculations at every $20$~fm/c.
For the calculation of the ion-ion separation distance $R$ we use the hybrid method, which
relates the coordinate to the quadrupole moment for small $R$ values, as described in
Ref.~\cite{UO09b}. The accuracy of the density constraint calculations is
commensurate with the accuracy of the static calculations.
\begin{figure}[!htb]
\includegraphics*[scale=0.38]{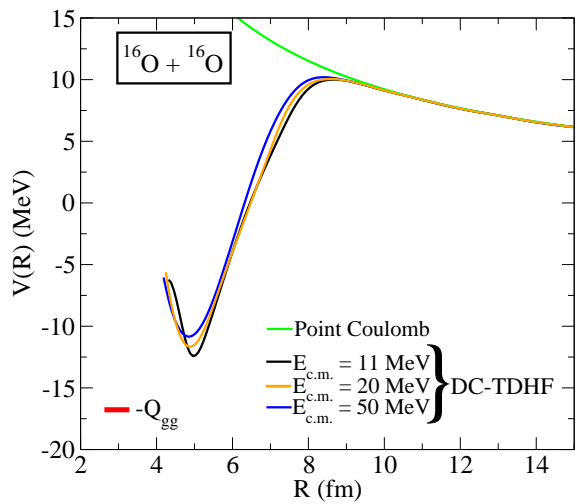}
\caption{Internuclear potential $V(R)$ for the head-on collision of the $^{16}$O+$^{16}$O system
for various $E_{\mathrm{c.m.}}$ values. The relative ground-state binding energy of the $^{32}$S nucleus
is represented by the $-Q_{gg}$ value.}
\label{fig:1}
\end{figure}

In order to develop a better insight into the excitation energy given by Eq.~(\ref{eq:1})
we have first studied two spherical systems, $^{16}$O+$^{16}$O and $^{40}$Ca+$^{40}$Ca.
In literature  one commonly  defines the excitation energy for a particular reaction as
\begin{equation}
 E^{*}=E_{c.m.}+Q_{gg}\;,
\label{eq:esgg}
\end{equation}
where  $Q_{gg}$ is the mass difference between the two initial nuclei
and the combined system in its ground state. While this expression is
correct relative to the ground state of the composite system, it does not
accurately describe the excitation energy relative to other intermediate
transition states formed during the collision.
Our choice of the reactions mentioned above was motivated by the fact that
the former system has a positive $Q_{gg}$ value
(16.6~MeV), whereas the latter system has a negative one (-14.2~MeV).

In Fig.~\ref{fig:1} we show the ion-ion interaction potential $V(R)$
for the head-on (zero impact parameter) collision of the $^{16}$O+$^{16}$O
system at various center-of-mass energies.  These results are essentially
the same as those published in Ref.~\cite{UO06b} except for the energy
dependence of $V(R)$. This dependence arises from the time available
for the system to undergo rearrangements and partial equilibration,
which approaches the frozen-density limit at high energies~\cite{WL09}.
\begin{figure}[!htb]
\includegraphics*[scale=0.38]{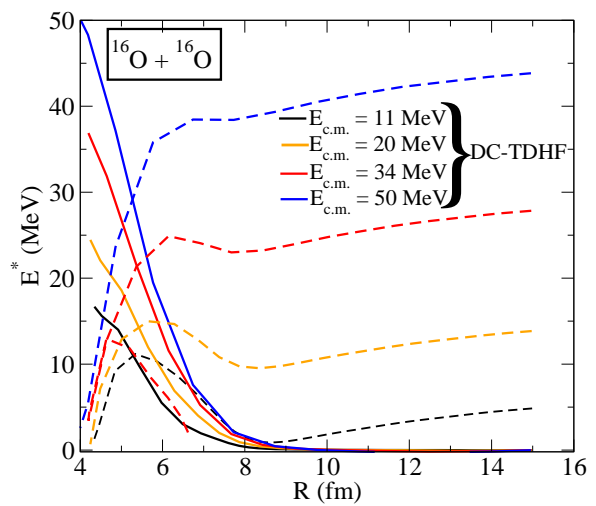}
\caption{Excitation energy for the head-on collision of the $^{16}$O+$^{16}$O system
for various $E_{\mathrm{c.m.}}$ values (solid curves) as a function of the ion-ion
distance $R$.  Also, shown are the corresponding collective kinetic energy $E_{kin}$ values (dashed curves).}
\label{fig:2}
\end{figure}
On the same figure we have also shown the relative location of the $^{32}$S ground-state 
binding energy as represented by the $-Q_{gg}$ value.
In Fig.~\ref{fig:2} the corresponding excitation energies are shown as a
function of $R$ calculated from Eq.~(\ref{eq:1}).
The fact that the excitation energy should be asymptotically zero is
a good test of numerical accuracy, which is reproduced quite well as
can been seen from the figure.
\begin{figure}[!htb]
\includegraphics*[scale=0.38]{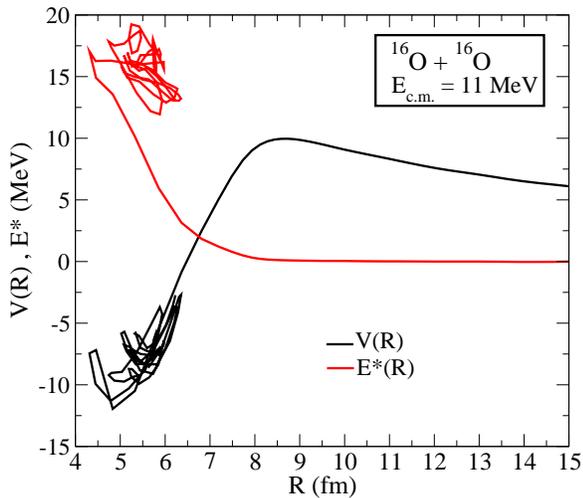}
\caption{Long-time evolution of the excitation energy, $E^{*}$, and the ion-ion potential, $V(R)$,
for the head-on collision of the $^{16}$O+$^{16}$O system at
$E_{\mathrm{c.m.}}=11$~MeV as a function of the ion-ion distance $R$.}
\label{fig:3}
\end{figure}
The final value of $E^{*}$ and smallest $R$ value are
chosen to be the ones corresponding to the smallest relative velocity
or smallest collective kinetic energy in the vicinity of the potential minimum,
which alternately can be referred to as the capture point.
Naturally, some of these quantities can only be pinned down within the
limits of density-constraint frequency.
For sake of completeness in Fig.~\ref{fig:3} we also show the
long-time behavior of the potential and the excitation energy for the
$^{16}$O+$^{16}$O system at $E_{\mathrm{c.m.}}=11$~MeV.
The figure demonstrates very nicely that the majority of the entrance energy is
absorbed into intrinsic motion such that the compound stage is stuck
in the vicinity of a certain $R$ value with rather small oscillations.
In the entrance phase of the collision, the excitation energy
$E^*$ increases monotonically with decreasing $R$.  After the point of
closest approach has been reached, the system is stuck close to that distance and the
dynamical evolution turns abruptly to irregular oscillations in $E^*$
as seen in Fig.~\ref{fig:3}, which the typical energy fluctuations of an excited ensemble.
\begin{figure}[!htb]
\includegraphics*[scale=0.38]{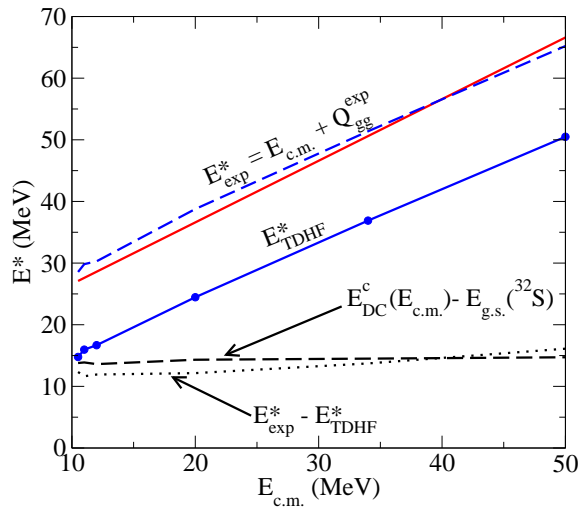}
\caption{Excitation energy for the head-on collision of the $^{16}$O+$^{16}$O system
for various $E_{\mathrm{c.m.}}$ values at the point of capture (solid blue line) and the
excitation energy calculated from Eq.~(\ref{eq:esgg}) (solid red curve). Other curves
are described in the manuscript.}
\label{fig:4}
\end{figure}
These fluctuations are here rather large
due to the small particle number. The statistical estimate for the
energy width is $\Delta E\approx\sqrt{16\varepsilon_\mathrm{F}E^*/(N\pi^2)}\approx 6$~MeV,
which fits nicely to the observed fluctuations.

Next we wanted to demonstrate the conjecture that the value of the excitation
energy measured at the capture point in TDHF is the excitation relative to
the composite or dinuclear system formed during the collision. In Fig.~\ref{fig:4}
we plot the excitation energy as a function of the center-of-mass energy for
both the analytic expression of Eq.~(\ref{eq:esgg}) (solid red line) and the
TDHF results at the capture point as discussed above (solid blue line).
As expected the TDHF result is below the one generated from Eq.~(\ref{eq:esgg}).
The difference between the two curves is shown by the dotted line. 
We have also calculated the energy difference of the composite system relative to the
ground state, which is simply the 
$E_{DC}$ at the point of capture minus the ground state energy of the $^{32}$S
system obtained by an unconstrained Hartree-Fock calculation.
\begin{figure}[!htb]
\includegraphics*[scale=0.38]{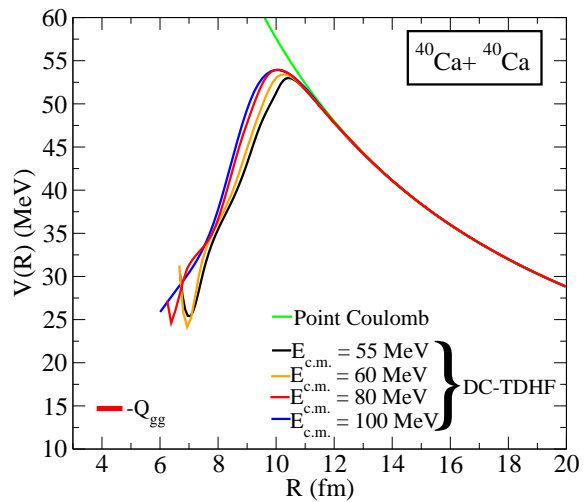}
\caption{Internuclear potential  for the head-on collision of the $^{40}$Ca+$^{40}$Ca system
for various $E_{\mathrm{c.m.}}$ values. The relative ground state binding energy of the $^{80}$Zr nucleus
is represented by the $-Q_{gg}$ value.}
\label{fig:5}
\end{figure}
This is shown by the black dashed curve. If we shift the TDHF result by these
differences we obtain the blue dashed curve, which is in agreement with the
result obtained from Eq.~(\ref{eq:esgg}). We should emphasize at this point
that for many reactions the excitation energy at the capture point is of great
interest as opposed to the ground state to ground state value, since after the
capture point many different reaction possibilities exist.

\begin{figure}[!htb]
\includegraphics*[scale=0.38]{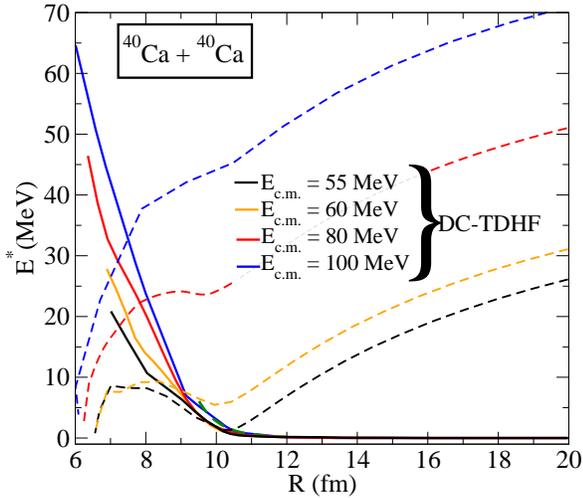}
\caption{Excitation energy for the head-on collision of the $^{40}$Ca+$^{40}$Ca system
for various $E_{\mathrm{c.m.}}$ values (solid curves) as a function of the ion-ion
distance $R$.  Also, shown are the corresponding collective kinetic energy $E_{kin}$ values (dashed curves).}
\label{fig:6}
\end{figure}

\begin{figure}[!htb]
\includegraphics*[scale=0.38]{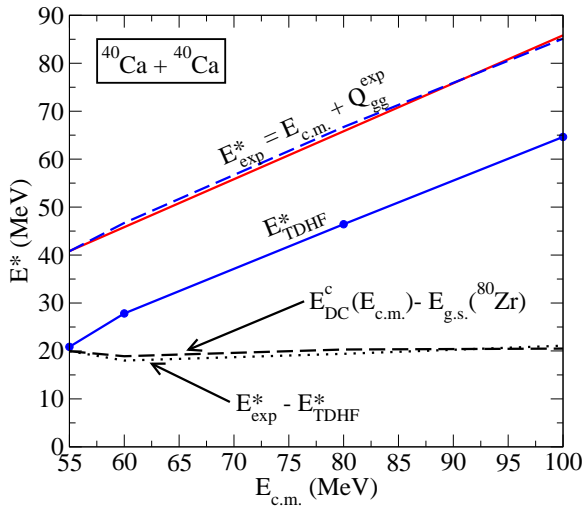}
\caption{Excitation energy for the head-on collision of the $^{40}$Ca+$^{40}$Ca system
for various $E_{\mathrm{c.m.}}$ values at the point of capture (solid blue line) and the
excitation energy calculated from Eq.~(\ref{eq:esgg}) (solid red curve). Other curves
are described in the manuscript.}
\label{fig:7}
\end{figure}

We have repeated the above study for the $^{40}$Ca+$^{40}$Ca system, for which the
$Q_{gg}$ value is -14.2~MeV. In Fig.~\ref{fig:5} we show the ion-ion potentials
calculated using the DC-TDHF method at a set of center-of-mass energies. The energy
dependence is analogous to the $^{16}$O+$^{16}$O case. 
Fig.~\ref{fig:6} shows the corresponding excitation energies calculated via TDHF
using Eq.~(\ref{eq:1}). Again, the excitation energies gradually rise as the nuclei enter
the interaction regime while the collective kinetic energies show a rise when the nuclei
first experience the nuclear attraction but eventually fall due to slowdown of the
composite system.
Finally, we again plot the center-of-mass energy dependence of the excitation energy at the
point of capture in Fig.~\ref{fig:7} (blue solid curve) together with the one obtained from
Eq.~(\ref{eq:esgg}), except this time using the negative $Q_{gg}$ value (solid red line).
Again the two curves run parallel to each other and their difference is shown by the
dotted line. The long dashed line shows the difference in energy between the density
constrained energy $E_{DC}$ at the point of capture minus the ground state energy of the $^{80}$Zr
nucleus. The two curves are almost exactly the same and shifting the TDHF excitation curve by this
energy difference produces the blue dashed curve, which is in excellent agreement with the
one obtained using Eq.~(\ref{eq:esgg}).
Thus, we can safely say that the microscopically calculated excitation energy represents the
excitation energy at the point of capture.
\begin{figure}[!htb]
\includegraphics*[scale=0.38]{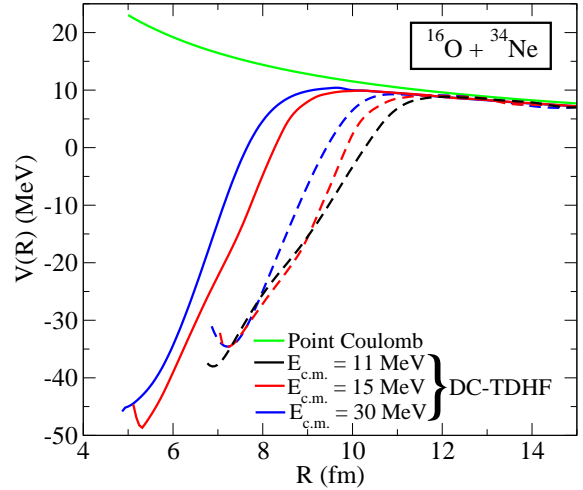}
\caption{Internuclear potential  for the head-on collision of the $^{16}$O+$^{34}$Ne system
for various $E_{\mathrm{c.m.}}$ values and two alignments of the $^{34}$Ne nucleus.
The solid lines denote the potential for vertical alignment whereas the dashed curves are
for the horizontal alignment of the $^{34}$Ne nucleus with respect to the collision axis.}
\label{fig:8}
\end{figure}

It is also possible to obtain an approximate temperature by relating the excitation energy
to the temperature using the Fermi gas model, $E^{*}=aT^2$,
where $a\approx A/8$~MeV$^{-1}$ is the level density parameter.
For the above reactions and the $E_{\mathrm{c.m.}}$ values used,
this translates into a temperature range of
$2.0-3.5$~MeV for the $^{16}$O+$^{16}$O system and a
temperature range of $1.4-2.5$~MeV for the $^{40}$Ca+$^{40}$Ca
system. The reliability of the above approximation should be higher
for heavier systems.

In order to examine how deformation influences the excitation energy during
a heavy-ion collision we have chosen to investigate the $^{16}$O+$^{34}$Ne
system. In Hartree-Fock calculations the neutron-rich $^{34}$Ne nucleus has
a large axially symmetric deformation. In the past we have examined the
effects of deformation on the ion-ion potentials due to the different
initial alignments of the deformed nucleus~\cite{UO06b,UO06e}.
We have performed TDHF collisions of $^{16}$O+$^{34}$Ne at various
center-of-mass energies and for the two extreme alignments of the
$^{34}$Ne nucleus, one in which the elongation axis is along the
collision axis and the other for which it is perpendicular.
In Fig.~\ref{fig:8} we show the ion-ion potentials obtained for this
system using the DC-TDHF method.
We observe that at $E_{\mathrm{c.m.}}=11$~MeV the vertical alignment case does not fuse,
whereas the horizontal alignment does. Furthermore, the centroids of the potentials are
different for the two alignments since the nuclei start to come into contact earlier/later
for horizontal/vertical alignments. Otherwise the energy dependence of the potentials is
commensurate with the spherical systems.
The fact that different orientations of the deformed nucleus lead to a difference in the
time of contact between the target and the projectile is expected to manifest itself in
the evolution of the excitation energy as well.
\begin{figure}[!htb]
\includegraphics*[scale=0.38]{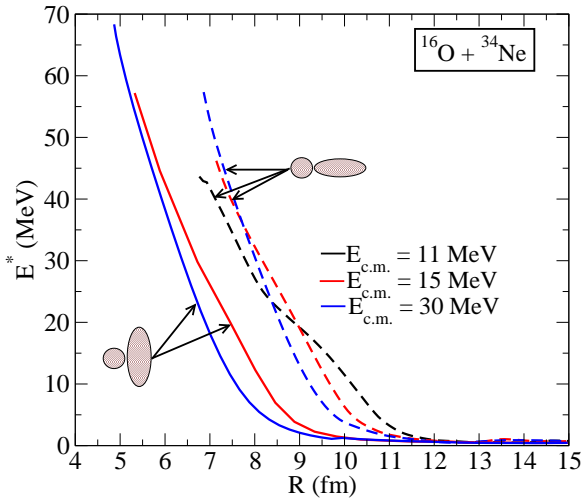}
\caption{Excitation energy for the head-on collision of the $^{16}$O+$^{34}$Ne system
for various $E_{\mathrm{c.m.}}$ values as a function of the ion-ion
distance $R$.
The solid lines denote the excitation energy for vertical alignment whereas the dashed curves are
for the horizontal alignment of the $^{34}$Ne nucleus with respect to the collision axis.}
\label{fig:9}
\end{figure}
Indeed, in Fig.~\ref{fig:9} we see that the excitation for the horizontal alignment of
the $^{34}$Ne nucleus starts earlier but somehow does not reach as large a value as the
vertical alignment case.
This is a very interesting observation since it would indicate that due to the differences
in the excitation energy at the point of capture different alignments will likely have
different probabilities for various exit channels.
For example, this could be a very important consideration for superheavy formations.

In summary, we have outlined a microscopic approach for calculating excitation energies of composite
or dinuclear systems formed during heavy-ion collisions.
The goal of the approach is to provide estimates for excitation energies at the
initial point of capture, after which a multitude of exit-channel possibilities may
exist and will be strongly influenced by the amount of excitation.
The premise of our approach depends on the reliability of TDHF theory for
describing the early stages of heavy-ion reactions.
As discussed in the manuscript, there is mounting evidence that TDHF does
provide a reliable description of heavy-ion dynamics and dissipation in this
stage whereas the long-time evolution may be questionable.
In order to elucidate the above arguments we have performed a number of
calculations involving both spherical and deformed nuclei.
It is our long-term goal to extend these calculations to heavier systems and in
particular to superheavy nuclei.

This work has been supported by the U.S. Department of Energy under grant No.
DE-FG02-96ER40963 with Vanderbilt University, and by the German BMBF
under contracts Nos. 06FY159D and 06ER142D.

\end{document}